\documentclass[10pt, oneside]{article}   	
\usepackage{geometry}                		
\usepackage{graphicx}				
\usepackage{amssymb,hyperref,multicol}
\hypersetup{colorlinks=true, citecolor=red, linkcolor=blue}
\usepackage{comment,color}

\usepackage{tocloft}
\usepackage[tocflat]{tocstyle}

\title{Joint analysis of Borexino and SNO solar neutrino data and reconstruction of the survival probability}
\author{Francesco Vissani\\[1.4ex]
\small INFN - Laboratori Nazionali del Gran Sasso, 
Assergi (AQ), Italy\\
\small Gran Sasso Science Institute, L'Aquila (AQ), Italy}
\date{}                                           

\begin{document}

\parskip=0.7ex
\maketitle

\begin{abstract}
Solar neutrino oscillations are supported by KamLAND's antineutrino measurements, but certain solar neutrino data--the observed shape of the ${}^8$B flux and the difference between day and night counting rates measured in Super-K--do not fit well with the ensuing oscillation pattern.
Interestingly,  other solar neutrino data allow independent tests of the survival probability.
Thanks to the new measurements of Borexino 
at low-energies along with the standard solar model and to the results of SNO at high-energies, four values of the neutrino survival probability are known. We  build and study a likelihood based only on these solar neutrino data.  The results
agree well with the standard oscillation pattern and in particular with KamLAND findings. A related and straightforward procedure permits to reconstruct the survival probability of solar neutrinos and to assess its uncertainties, for all solar neutrino energies. 
\end{abstract}




\section{Introduction}
Solar neutrinos continue to provide valuable occasions of research
to experimentalists and theorists working  in astrophysics and  
in particle physics. In the present work, we aim at a fresh assessment of solar neutrino oscillations and at reconstructing the survival probability, by    exploiting the new experimental results made available by Borexino Collaboration~\cite{borex}. 
In the rest of this section, we describe in greater detail the underlying context and motivations.

The MSW theory of neutrino oscillations~\cite{msw} is widely considered reliable and consistent with other facts,
e.g., with SNO neutral current results~\cite{Bellerive:2016byv} and with 
KamLAND terrestrial antineutrino measurements~\cite{kl}.
However, in a recent paper of Super-Kamiokande Collaboration one reads that~\cite{Abe:2016nxk},
\begin{quote}
{\em there is still no clear evidence that the solar neutrino flavor conversion is indeed due to neutrino oscillations and not caused by another mechanism.}
\end{quote}
Indeed, the measurements of ${}^8$B neutrinos of Super-Kamiokande~\cite{Abe:2016nxk} and SNO~\cite{Bellerive:2016byv} 
are consistent with a constant suppression of the expected flux. 
Super-Kamiokande Collaboration~\cite{Abe:2016nxk} finds a hint for day-night effect but 
does not see evidence of a decrease with the energy of the survival probability (in solar neutrino jargon,  
 ``upturn'' means commonly a negative and measurable value of the slope at ${}^8$B energies).
SNO results~\cite{Bellerive:2016byv} do not contradict these results even if they are less significant.
These results favor values of the oscillation parameter 
$\Delta m^2_{21}$ that are smaller 
and 2$\sigma$ away from those pointed out by the global analyses, see e.g.~\cite{Capozzi:2017ipn}, \cite{concha},  \cite{jose}, 
that are mostly due to KamLAND {and not by solar neutrino data themselves}.  
More solar neutrino data are necessary to settle the issue.

{In fact,  this situation has stimulated the theoretical debate and 
new physics scenarios have been proposed, see e.g., \cite{dighe,palazzo},  
and also \cite{Vissani:2017dto} for a recent assessment.}

Here, we extract the  
parameters of MSW theory using other and independent solar neutrino data. 
We include in the analysis the counting rates of  4 different branches of the solar neutrinos, 
measured by SNO and Borexino, the latter just appeared and not yet used in global analyses.  
The KamLAND results, the results on  day-night asymmetry, the 
spectral shape of the ${}^8$B neutrinos, will not be used instead. 
In this manner, the results of our {\em analysis of solar neutrino data} can be compared with the other ones, 
verifying the consistency; as we will see, the results agree very well with the global fits 
and in particular with KamLAND.\footnote{The results of 
Homestake, Gallex/GNO and SAGE are relevant for the current 
global fits. However, we do not include these 
integral measurements (=that sum the contributions of various solar neutrino 
branches) since they cannot be directly attributed to a specific energy.} 
We show how to use this type of analysis to  reconstruct the survival probability 
quite precisely.

\section{The MSW survival probability}

The survival probability of electron neutrinos from the Sun, that includes three flavor effects, can be conveniently
approximated as, 
\begin{equation}
P 
(E_\nu\ ; \Delta m^2_{21},\theta_{12})=\cos^4\!\theta_{13}\times P_{\mbox{\tiny 2f}}(E_\nu\ ; \Delta m^2_{21},\theta_{12}) + 
\sin^4\!\theta_{13}
\end{equation}
where $\sin^2\theta_{13}\approx 0.022$ is well-known and will be kept fixed in the analysis. 
To simplify the notation of the theoretical (true)
survival probability $P$, we do not use any superscript or subscript;   
instead, we will use a subscript to identify 
the experimental values $P_i$,  discussed just below.
This probability depends slightly upon whether the neutrinos are detected on day or on night;
however, for the region of parameters in which we are interested, 
this effect at most \%, and it is much smaller for energies below the ones of ${}^8$B neutrinos. 
We  will 
consider the average survival probability between day and night,
\begin{equation}
P_{\mbox{\tiny 2f}}=\frac{P_{\mbox{\tiny 2f}}^{\mbox{\tiny day}}+ P_{\mbox{\tiny 2f}}^{\mbox{\tiny night}}}{2}
\end{equation}
and we evaluate the theoretical expression of the probability at a fixed energy.
The standard two-flavor formulae are adopted, namely, 
$P_{\mbox{\tiny 2f}}^{\mbox{\tiny day}}
=\frac{1}{2} \left( 1+ \cos 2\theta_{12} \times \cos 2\theta_{12}^{m} \right)$
and 
$P_{\mbox{\tiny 2f}}^{\mbox{\tiny night}}=
P_{\mbox{\tiny 2f}}^{\mbox{\tiny day}} + 
\mbox{reg}( E_\nu\ ; \Delta m^2_{21},\theta_{12})$; 
the matter mixing angle  $\cos 2\theta_{12}^{m}( E_\nu\ ; \Delta m^2_{21},\theta_{12})$ and 
the regeneration function $\mbox{reg}( E_\nu\ ; \Delta m^2_{21},\theta_{12})$ are evaluated 
analytically with the expressions summarized in~\cite{Vissani:2017dto}.
We do not separate day and night data and we do not use the information on the 
shape of the ${}^8$B  neutrino for our analysis: stated otherwise, 
the hint for day-night asymmetry and the (lack of) upturn at lower energies will be 
regarded as independent data, that lead to independent conclusions.

\begin{figure}[t]
\begin{minipage}[c]{4.7cm}
\caption{\em\small 
Two-flavor survival probabilities curves, from~\cite{Vissani:2017dto}, 
for neutrinos produced in the center of the Sun and shown separately 
for those that arrive by day and those that arrive by night.  \medskip\newline
The superimposed, colored arrows indicate the 
energies of the pp-branches directly observed by Borexino and SNO.
 \label{f0}}
 \end{minipage}
 \begin{minipage}[c]{10.5cm}
\includegraphics[width=10.5cm, trim=-20mm 0mm 0mm 0mm]{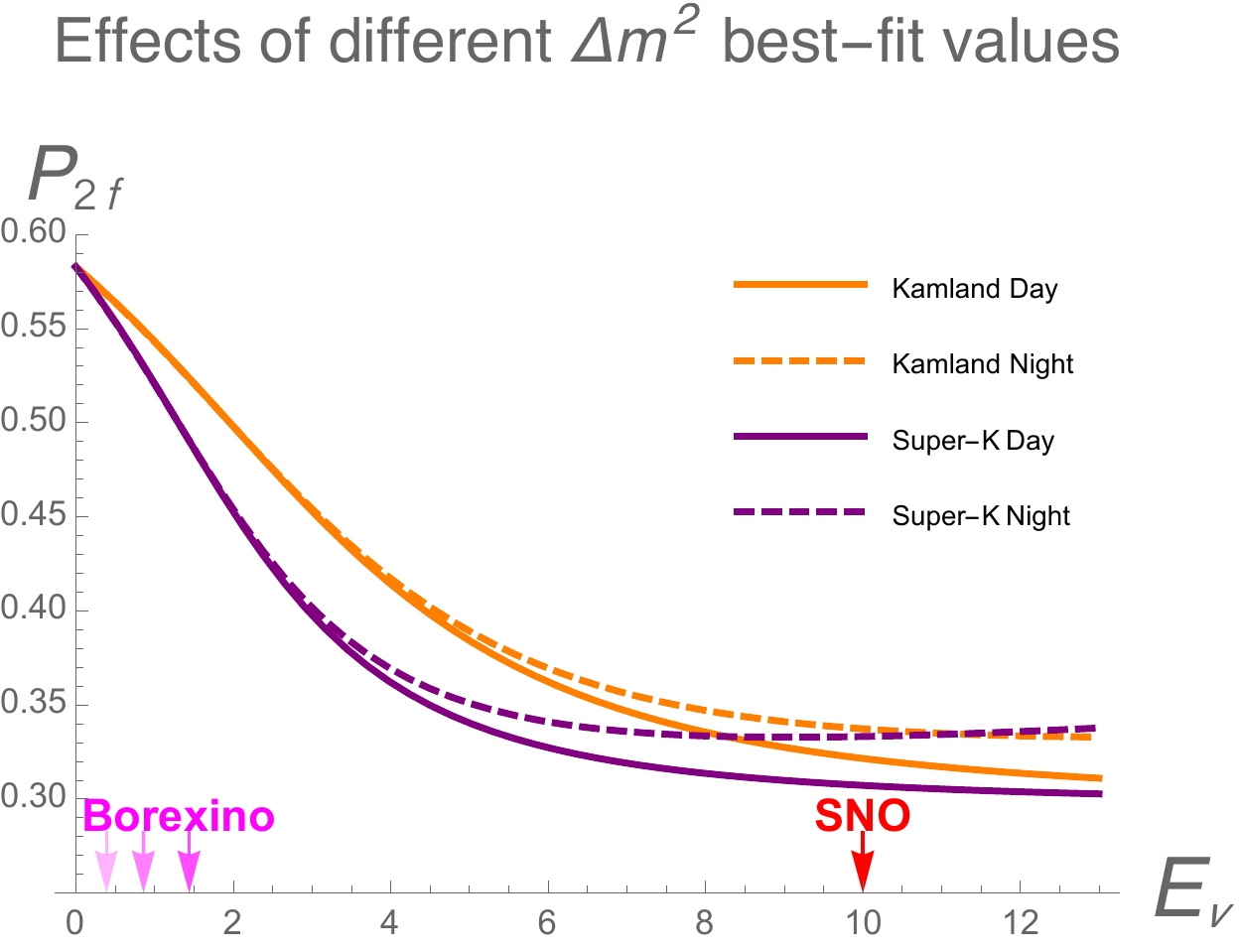}
\end{minipage}
\end{figure}

\paragraph{Illustration}
To summarize and for the purpose of illustration, we show in Fig.~\ref{f0} the two-flavor survival probabilities 
given in~\cite{Vissani:2017dto} for parameters consistent with the KamLAND observations, $\Delta m^2_{21}=7.4\times 10^{-5}$ eV$^2$ and with Super-Kamiokande observations alone, $\Delta m^2_{21}=4.9\times 10^{-5}$ eV$^2$. 
When $\Delta m^2_{21}$ decreases, the ${}^8$B neutrinos enter  deeper into the MSW region and get further away from the transition region; thus, at ${}^8$B energies,  both regeneration effect and the slope of the curve increase. 

Also another (evident) effect occurs: {\em The values of the survival probability  at other energies change.}
In the present work, we will exploit this remark to determine the parameters of solar neutrino oscillations and to reconstruct the shape of the survival probability, since, as a matter of fact, solar neutrinos have been observed at `high-energy', $\sim 10\ \mathrm{MeV}$ (${}^8$B neutrinos) 
but also at  `low-energy' region, $\sim 1\ \mathrm{MeV}$ ($pp$, ${}^7$Be, $pep$ neutrinos), 
that has been emphasized in Fig.~\ref{f0}.

{\paragraph{Low energy behavior of the various survival probabilities \label{lj}}
The various type of neutrinos are produced in somewhat different regions of the Sun; 
this fact matters 
for an accurate description of  the survival probabilities of electron neutrinos. 
In order to show the point most clearly, we consider the low energy regime, $E_\nu\sim 1$ MeV,  
when the corrections due to MSW theory are small and the regeneration function (due to terrestrial matter density)
is even smaller. 
In this regime, and using the notations of~\cite{Vissani:2017dto},
we note that there is a small parameter,
\begin{equation}
 \epsilon_\odot\approx 1.04 \left( \frac{n_e^\odot}{100\; \mbox{mol}} \right)\times 
 \left( \frac{7.37\times 10^{-5}\mbox{ eV}^2}{\Delta m^2_{21}} \right)\times 
 \left( \frac{E_\nu}{5\; \mbox{MeV}} \right)
\end{equation}
Then we can then Taylor-expand the oscillations probabilities in 
$\epsilon_\odot$, finding,  
\begin{equation}
P_{\mbox{\tiny 2f}}^{\mbox{\tiny day}}=\left(1-\frac{\sin^2 2\theta_{12}}{2} \right)
-\frac{\cos2\theta_{12}\; \sin^2 2\theta_{12} }{2} \times \epsilon_\odot +\mathcal{O}( \epsilon_\odot^2)
\label{spb}
\end{equation}
where the first term is the usual expression of the vacuum survival probability, and 
the second one is contributed by the MSW theory. 
Averaging Eq.~(\ref{spb}) over the region of neutrino production, the  electron densities get replaced by its average
values $\langle n_{e,i}^\odot\rangle $, that are different for different neutrino 
species $i$.
Their values can be calculated by means of the standard solar model. Using~\cite{bahcall} and adopting the version with  
OP opacities we find,
\begin{equation}
\langle n_{e,i}^\odot\rangle =61.8,\ 67.8,\ 81.1,\ 89.6\ e^- \mbox{ mol} \mbox{ for  }i=pp,\ pep,\ {}^7\mbox{Be},\ {}^8\mbox{B}
\label{bave}
 \end{equation}
 (To be sure, the neutrinos that are produced in deeper regions have a bigger $\langle n_e^\odot\rangle$;
 thus, their survival probability
 decreases a bit faster with increasing energies.) 
 
For example, the survival probability of $pp$ neutrinos can be
 approximated as,
\begin{equation}
P_{\mbox{\tiny 2f}}^{\mbox{\tiny day}}=0.583 \left( 1 - 1.6\% \times \frac{E_\nu}{0.42\mbox{ MeV}}  \right)
 \end{equation} 
 where, for this numerical evaluation, we use the oscillation parameters at the current best fit point.
 The corresponding distortion of the $pp$ neutrino spectrum 
 is small but potentially interesting for future, very precise measurements.
 
Two last remarks are in order: 
1)~the survival probability that is usually shown and discussed is the one that concerns 
${}^8$B neutrinos, namely, those that have been studied by Super-Kamiokande and SNO.
2)~The survival probabilities, measured using $pp$, $pep$ or ${}^7$Be
 neutrinos at certain energies, {\em are slightly larger} than the  survival probability of ${}^8$B 
 neutrinos measured
 at the same energies. This is conceptually interesting even if the difference is
 just percent (as will be quantified more precisely later).
These remarks will be relevant, in particular, for the discussion of Sect.~\ref{sec:rec}.}

\section{Expectations for the solar neutrino fluxes\label{cagoi}}

In order to investigate the transformations of solar neutrinos, and in particular those attributable to the 
MSW theory discussed just above, expectations on the fluxes
{\em before} neutrino oscillations are necessary. 
The basic tool for this purpose is the standard solar model (SSM) originally developed by J.~Bahcall more than half a century ago, tested and improved  in the course of the years by him and by many collaborators.
The residual uncertainties of the model depend upon several factors, 
including nuclear physics, opacities, and solar abundances of $Z>2$ elements (`metallicity').
Among the ways to validate the SSM, the main one 
to date remains the observation and interpretation of helioseismic p-modes. 

The most recent and accurate version of the SSM to date is documented in~\cite{Vinyoles:2016djt}. The two models for solar abundances used there 
lead to significantly different predictions for helioseismic observations: The one  with higher metallicity, called there B16-GS98,
compares reasonably well with {these observations}; the other one with low metallicity does not~\cite{Vinyoles:2016djt}. 
For this reason, we adopt the expectations of the former version of the SSM for 
the $pp$ and for the $pep$-neutrinos,  
$\Phi^{\odot}_{\mbox{\tiny pp}}=\Phi^{\mbox{\tiny B16-GS98}}_{\mbox{\tiny pp}}$ 
and $\Phi^{\odot}_{\mbox{\tiny pep}}=\Phi^{\mbox{\tiny B16-GS98}}_{\mbox{\tiny pep}}$.

For what concerns the flux of ${}^8$B neutrinos, it is important to emphasize that this has been measured directly by SNO experiments with neutral current reactions~\cite{Bellerive:2016byv}. 
The determination of SNO is compatible but more precise than the one that is  
provided us by the current version of the  SSM, as can be seen by the following comparison,
\begin{equation}
\Phi^{\mbox{\tiny B16-GS98}}_{\mbox{\tiny B}}=5.46 (1\pm 0.12) \times 10^6 /\mbox{cm}^{2}\mbox{s}, \ \ \ \  
\Phi^{\mbox{\tiny SNO}}_{\mbox{\tiny B}}=5.25 (1\pm 0.04) \times 10^6 /\mbox{cm}^{2}\mbox{s}
\label{corneto}
\end{equation}
It is convenient to use the flux of ${}^8$B neutrino observed by SNO, rather than the theoretical 
SSM prediction:  $\Phi^{\odot}_{\mbox{\tiny B}}=\Phi^{\mbox{\tiny SNO}}_{\mbox{\tiny B}}$. 
The key point is just that this expectation 
has nothing to do with oscillations (to be tested later); let us repeat that its 
advantage is that it implies an uncertainty smaller than the one of SSM.

Finally, we discuss the ${}^7$Be neutrinos. 
{Their flux depends upon the production reaction 
${}^3\mbox{He}+{}^4\mbox{He}\to {}^7\mbox{Be}+\gamma$, whose cross section 
is proportional and it is expressed in terms of the parameter $S_{34}$
(i.e., the S-factor, see e.g., \cite{wiss} for the definition).} 
Several extrapolations to solar energies of the available data on  ${}^3\mbox{He}+{}^4\mbox{He}\to {}^7\mbox{Be}+\gamma$ are present in the literature e.g.~\cite{Adelberg}, model based, and~\cite{deBoer}  based on $R$-matrix.
The values of $S_{34}$ obtained~\cite{Adelberg} and~\cite{deBoer} are consistent within uncertainties. However, in~\cite{deBoer}, 
the experimental data, including elastic scattering phase shifts, are consistently described over a wider energy range. Thus at the present status of knowledge, the determination of $S_{34}$ obtained in~\cite{deBoer} can be presumed to be more robust, as argued there  and further explained in~\cite{dileva}. Therefore, this will be used in the following.
 This implies a slight change of certain SSM fluxes and more precisely a 
 downward shift of the ${}^8$B and of the ${}^7$Be  
fluxes by 2.7\% and 2.8\% respectively~\cite{Vinyoles:2016djt};
note incidentally that 
this improves even further the agreement of the central values in Eq.~(\ref{corneto}).
 In view of these considerations, we
 will apply the predicted 2.8\% downward renormalization of the ${}^7$Be flux; on top of that, we assume conservatively the same error\footnote{It is plausible that, 
 if the predictions of  the SSM are enhanced by anchoring the ${}^8$B flux to the value measured by SNO, 
 this will have also other impacts on the predictions, 
 including a reduction of the uncertainty of ${}^7$Be flux; however, this goes beyond the scope of the present discussion.} of the high metallicity model, 
  \begin{equation}
  \Phi^{\odot}_{\mbox{\tiny Be}}=\Phi^{\mbox{\tiny B16-GS98}}_{\mbox{\tiny Be}} (1-0.028)
  \end{equation}
a small revision (improvement) of the nominal value of the flux from the  
B16-GS98 version of the SSM, that is largely within theoretical errors. 
Note that by adopting this smaller value leads to {\em increase} the value of the 
survival probability  measured by Borexino  by the same factor $1/(1-0.028)$, see next section.

Summarizing, the expectations for the solar neutrino fluxes that we adopt are,
\begin{equation}
\begin{array}{cc}
\Phi^{\odot}_{\mbox{\tiny pp}}=(5.98\pm 0.04)\times 10^{10} /\mbox{cm}^{2}\mbox{s}, & 
\Phi^{\odot}_{\mbox{\tiny Be}}=(4.79\pm 0.29)\times 10^{9} /\mbox{cm}^{2}\mbox{s}, \\[1ex]
\Phi^{\odot}_{\mbox{\tiny pep}}=(1.44\pm 0.01)\times 10^{8} /\mbox{cm}^{2}\mbox{s}, &
\Phi^{\odot}_{\mbox{\tiny B}}=(5.25\pm 0.20)\times 10^6/\mbox{cm}^{2}\mbox{s}.
\end{array}
\end{equation}
where the numerical values of the fluxes of the 
the $pp$ and $pep$ neutrinos are from 
high metallicity SSM (version B16-GS98), as given 
table~5 of~\cite{Vinyoles:2016djt}, and the other two are discussed above.

\section{Known values of the survival probability}

A straightforward strategy to reconstruct the pattern of solar neutrino oscillations is to constrain the 
survival probability of electron neutrinos using the measurements that have been 
obtained at various energies.
 This approach is possible using the results of those detectors, capable to isolate the individual branches of solar neutrinos, 
i.e., to measure 
some parts of the differential neutrino spectrum. These are Kamiokande, Super-Kamiokande, SNO, KamLAND and Borexino. 

Kamiokande, Super-Kamiokande, SNO and Borexino measured the electronic neutrinos from the ${}^8$B branch;
Super-Kamiokande and SNO attained the highest precision. 
SNO (as discussed above) measured also the total flux of neutrinos and thanks 
to these measurements, the suppression of the flux of the electronic neutrinos from the ${}^8$B  has a special status: it is proved 
experimentally.  
Borexino and subsequently KamLAND measured neutrinos from the beryllium line, the former experiment attaining a great precision; 
finally, Borexino probed  also the $pp$ and the $pep$ branches.  Thus, there are 4 measurements at different energies. 

%
%

The formulae for the expected numbers of events (due to {neutral  and charged 
current} on deuterium at SNO and due to elastic scattering  in all detectors) 
help to clarify how it is possible to measure the (average) survival probability, knowing the SSM prediction for a certain 
flux $ \Phi^{\mbox{\tiny SSM}}_i$, when the individual contribution can be tagged experimentally.
These expressions are, 
\begin{equation}
\begin{array}{rl}
N^{\mbox{\tiny $\nu$D,nc}}_{\mbox{\tiny B}}=& \mathcal{N}_{\mbox{\tiny D}} T \int dE_\nu \  \epsilon^{\mbox{\tiny nc}}(E_\nu) \times 
 \Phi^{\mbox{\tiny SSM}}_{\mbox{\tiny B}} \times \sigma^{\mbox{\tiny nc}}(E_\nu)     \equiv  N_{\mbox{\tiny $\nu$D,nc}}^{\mbox{\tiny SSM}} \\[1.5ex]
N^{\mbox{\tiny $\nu$A,cc}}_{\mbox{\tiny B}}=&\mathcal{N}_{\mbox{\tiny D}} T \int dE_\nu \  \epsilon^{\mbox{\tiny cc}}(E_\nu) \times 
 \Phi^{\mbox{\tiny SSM}}_{\mbox{\tiny B}}\times P_{\mbox{\tiny B}} \times \sigma^{\mbox{\tiny cc}}(E_\nu)     \equiv P_{\mbox{\tiny B}} \times N_{\mbox{\tiny $\nu$D,cc}}^{\mbox{\tiny SSM}} \\[1.5ex]
N^{\mbox{\tiny ES}}_i=&\mathcal{N}_e T \int dE_\nu \  \epsilon^{\mbox{\tiny ES}}(E_\nu) \times 
 \Phi^{\mbox{\tiny SSM}}_i\times
\left[ P_i  \times  \sigma^{\mbox{\tiny $\nu_e$}}(E_\nu)   + (1-P_i)\times 
\sigma^{\mbox{\tiny $\nu_\mu$}}(E_\nu)  \right]  \\[1ex]
 \equiv & P_i \times ( N_{\mbox{\tiny $\nu_e$}}^{\mbox{\tiny SSM}} - N_{\mbox{\tiny $\nu_\mu$}}^{\mbox{\tiny SSM}})
+ N_{\mbox{\tiny $\nu_\mu$}}^{\mbox{\tiny SSM}} \ \ \ \ \mbox{ with }i=pp,\ {}^7\mbox{Be},\ pep   ,\ {}^8\mbox{B}
\end{array}
\end{equation}
where $\mathcal{N}_{\mbox{\tiny D}}$ (resp., $\mathcal{N}_{e}$) is  the number of deuterons 
(resp., of electrons), namely, of targets; 
$T$ is the time of measurement;
$\epsilon$ are the efficiency functions; $\sigma$ the cross sections.
We consider the values of $P_i$ averaged between day and night (assuming that the 
detector efficiency is constant). For neutrinos from electron capture (monochromatic), 
the energy is well known;  for neutrinos from continuous distributions, 
we consider the average energy of the distributions. 
E.g., for Super-Kamiokande and with a threshold of 4.5 MeV, the average is at 
$9$ MeV whereas for SNO the average energy is at 10 MeV~\cite{Bellerive:2016byv};
the widths are in both cases few MeV.


\paragraph{High energy:}

The ${}^8$B  neutrinos, measured again at SNO with charged currents, 
$\Phi^{\nu_e,\mbox{\tiny SNO}}_{\mbox{\tiny B}}=(1.735\pm 0.090)$ in units of $10^6$ cm$^{-2}$s$^{-1}$, 
combining phase I and phase II values~\cite{Bellerive:2016byv}, can be compared with the one measured by neutral currents. 
The ratio gives directly the value of the survival probability
\begin{equation}
P_{\mbox{\tiny B}}= \frac{\Phi^{\nu_e,\mbox{\tiny SNO}}_{\mbox{\tiny B}}}{\Phi^{\mbox{\tiny SNO}}_{\mbox{\tiny B}}}=0.33\pm 0.02
\end{equation}
This procedure is advantageous. The charged current and neutral current have similar cross sections and the 
measurements are obtained with the same detector, thus, 
one may expect that some systematics cancel in the ratio. 
This value is very precise, the relative uncertainty being just 6\%. 
Interestingly, the main error ($\sim 5\%$) comes from the charged current measurement. 
%
%
%

It is possible to validate this result as follows. 
The SNO collaboration has also obtained a fit of the day and night energy spectra; in view of our goals and of other considerations\footnote{The asymmetry 
$A_{ee}=2(P^{\mbox{\tiny night}}-P^{\mbox{\tiny day}})/(P^{\mbox{\tiny night}}+P^{\mbox{\tiny day}})$ 
shows a decreasing trend with the energy~\cite{Bellerive:2016byv}, 
while if it was due to regular three flavor neutrino oscillations, it should increase.},
we use this analysis only as a test. Consider the average values over energy 
at $E_{{\mbox{\tiny B}}}=10$~MeV as given in~\cite{Bellerive:2016byv}.  The 
probability of survival on day time  is  $c_0=0.317 $ and the average night-day  asymmetry is $a_0=0.046$. Thus,   
the average between day and night is $P_{{\mbox{\tiny B}}}=(P^{\mbox{\tiny night}}+P^{\mbox{\tiny day}})/2=
c_0/(1-a_0/2)=0.324\pm 0.020$ where the statistics (dominating) and systematics errors are included. This is consistent with the result derived above, that will be adopted for the following calculations. 



\paragraph{Low energy:} 
Three low energy branches of the $pp$ chain, namely
the beryllium  line at 862~keV, the fundamental $pp$ branch, and the tightly connected 
$pep$  line, have been all measured precisely by Borexino~\cite{borex}.
The intensity of this beryllium  line is known with a precision that is twice better than the SSM prediction;  
moreover, the observation has been confirmed by KamLAND. The $pp$ neutrinos, that are directly 
linked to the solar luminosity, are also measured, although with limited precision; the 
related $pep$  neutrino flux is also probed, and the measurement depends slightly 
upon uncertain details of the SSM~\cite{borex}.
The best values of the survival probabilities  are given directly in~\cite{borex}, using the 
{B16-GS98} version of the SSM~\cite{Vinyoles:2016djt} and including the uncertainties to the SSM. 
In view of the discussion of Sect.~\ref{cagoi}, the  
value of the survival probability for ${}^7$Be  neutrinos  cited in~\cite{borex},
will increased by  
$ P_{\mbox{\tiny B}}\to P_{\mbox{\tiny B}}/(1-0.028)$ while the values of 
$ P_{\mbox{\tiny pp}}$ and $ P_{\mbox{\tiny pep}}$ are just the same as in~\cite{borex}.

\bigskip
The known four values of the survival probability are summarized in Tab.~\ref{tab1}. Note that the uncertainties 
in the first three values 
include those of the SSM. 

\begin{table}[t]
\centerline{
\begin{tabular}{lcccc}
$i$-th solar   & source   & energy & known & dominant \\
branch &   & [MeV] & value of $P_i$       &  error  \\ \hline
$pp$ & {\footnotesize Borexino+SSM} & $\approx 0.39$ & $0.57\pm 0.10$ &  experiment \\
 ${}^7\mbox{Be}$ & {\footnotesize Borexino+SSM} & 0.862 & $0.545\pm 0.05$ &  theory \\
$pep$ &{\footnotesize Borexino+SSM} & 1.442  & $0.43\pm 0.11$ &  experiment \\ 
  ${}^8\mbox{B}$&{\footnotesize  SNO} & $\approx 10$ & $0.33\pm 0.02$ &  experiment
\end{tabular}}
\caption{\em\small  The four known values of the survival probabilities--see the text for a discussion. 
The first three values are collectively referred to as `low-energy' values whereas
the one corresponding to the ${}^8$B neutrinos is called `high-energy' value.\label{tab1}}
\end{table}



%
%
%
  

\section{Analysis of the oscillation parameters}

\begin{figure}[t]
\centerline{\includegraphics[width=0.95\textwidth]{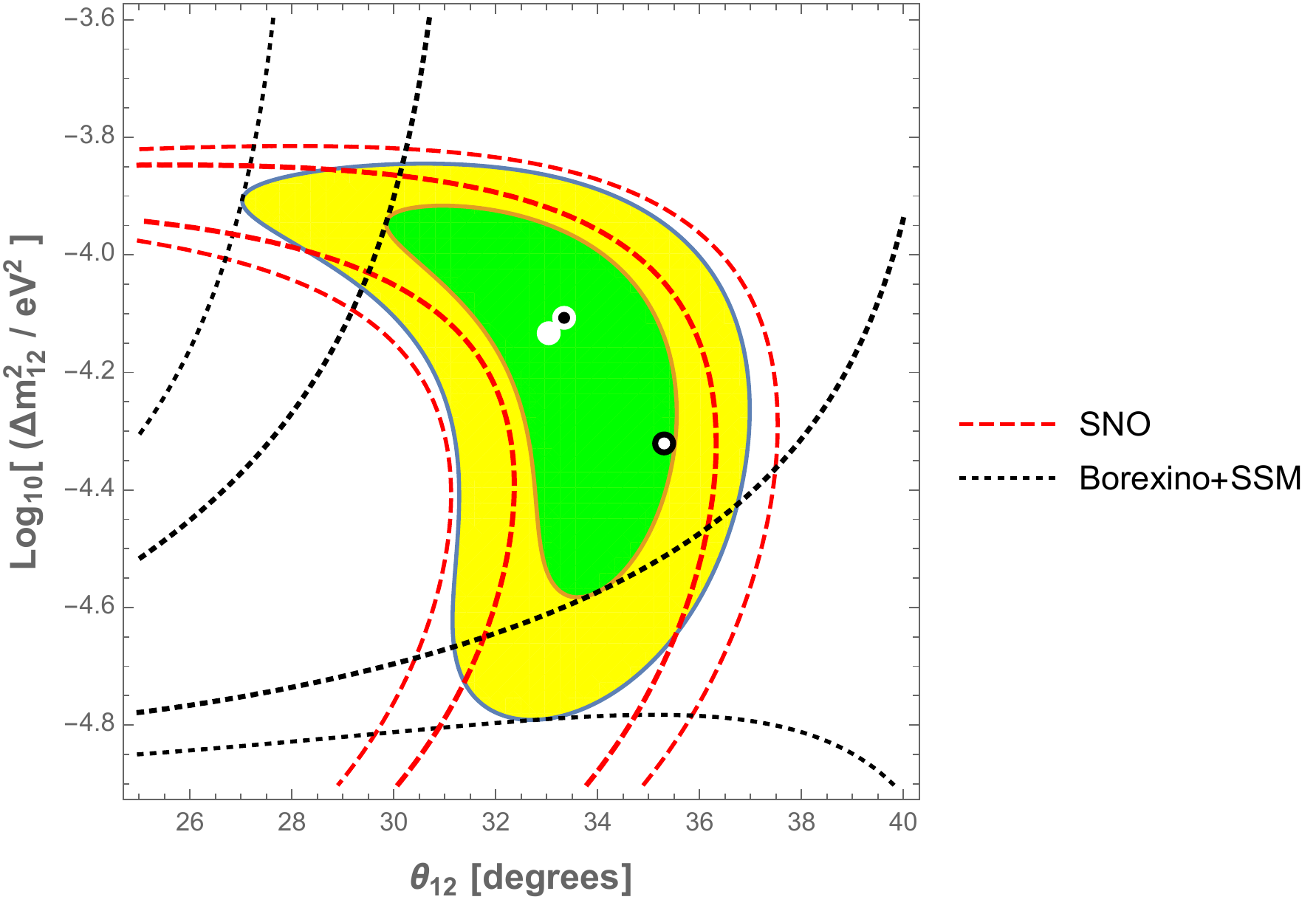}}
\caption{\em\small The two-dimensional areas filled in green (yellow) enclose the 68.3\% (95\%) confidence regions
of our solar neutrino analysis. We show  separately
the impact of the three values of the survival probability  known at low energies (Borexino+SSM,  dotted lines) and 
of the single value known at high energy (SNO, dashed lines). {The three circles show three best fit points, given in 
Eq.~(\ref{valori}):}
The best fit point of this analysis is given by the white disk dotted in black;
the white disk indicates the best global fit value; the black disk dotted in white is the 
best global fit of Super-Kamiokande data alone.
 \label{f1}}
\end{figure}

\paragraph{Method:}
The unambiguous measurements of the solar neutrino fluxes for 
several branches of the $pp$ chain, along with reliable theoretical SSM expectations, 
gives us four values of the survival probability $P_i\pm \delta P_i$. 
This allows us to adopt a very direct, chi-square based
procedure of analysis of the survival probability, 
\begin{equation}
\chi^2(\Delta m^2_{21},\theta_{12})=\sum_{i}\frac{\left(\ P(E_i\, ;\, \Delta m^2_{21},\theta_{12}) -P_i\ \right)^2 }{\delta P_i^2}
\end{equation}
A more complete notation for the true survival probability would be, 
$P(E_i\, ;\, \Delta m^2_{21},\theta_{12},\theta_{13}; n_{e,i}^\odot)$, 
but the mixing angle $\theta_{13}$ is measured very precisely by terrestrial experiments and therefore 
is kept fixed in this analysis 
and  
{likewise, the production densities of the neutrinos $n_{e,i}^\odot$ 
are assumed to be known precisely enough and are set to their average values,\footnote{This approximation for the average survival probability allows us to reduce the computational load; it implies an error of 0.3\% at 10 MeV, acceptable for our purposes and much better for energies around MeV (see Sect.~\ref{lj}).} 
given in Eq.~(\ref{bave}).}
The index $i$ runs over the types of neutrinos that are included in the analysis. It is possible to associate this 
chi-square to a likelihood in the usual manner, 
\begin{equation}
\mathcal{L}(\Delta m^2_{21},\theta_{12})\propto \exp\!\left[ - \frac{\chi^2(\Delta m^2_{21},\theta_{12})}{2} \right]
\end{equation}
that is normalized to unity in  the (prior) search window $10^{-5} \mbox{ eV}^2\le \Delta m^2_{21} \le 10^{-3} \mbox{ eV}^2$ 
and $20^\circ\le \theta_{12}\le 40^\circ$. 
The confidence levels of two-dimensional likelihood are analyzed by using a 
2-degrees-of-freedom Gaussian procedure 
\begin{equation}
\mathcal{L}(\Delta m^2_{21},\theta_{12})\ge  \mbox{(1-C.L.)}\times \mathcal{L}_{\mbox{\tiny best fit}}
\end{equation}

\paragraph{Results:}
The main result of the $\chi^2$ analysis is given in Fig.~\ref{f1}.
The returned best fit value of $\Delta m^2_{21}$ is rather close to
the one obtained in the global analysis~\cite{Capozzi:2017ipn} (consistent with~\cite{concha} and~\cite{jose})
{that is driven by KamLAND findings and not by solar data, while it is somehow larger than the value indicated by  Super-Kamiokande alone~\cite{Abe:2016nxk} that includes their analysis of the shape of ${}^8$B neutrinos 
and their measurement of the day-night asymmetry. These three values are,}
\begin{equation}
\begin{array}{c|c|c}
 \mbox{best fit}  & \log_{10}[\Delta m^2_{21}/\mbox{eV}^2] & \theta_{12} \\[0.5ex]  \hline 
 & & \\[-2ex]
    \mbox{{this work}}  \ \  &\ \  {- 4.11}  &\ \  {33.4^\circ} \ \ \\[0.5ex] 
 \mbox{global} \ \  &\ \  - 4.13  &\ \  33.0^\circ \ \ \\[0.5ex] 
  \mbox{Super-K only}   \ \  &\ \  - 4.32  &\ \  35.0^\circ \ \ 
\end{array} \label{valori}
\end{equation}
These values are displayed in Fig.~\ref{f1}; 
note that all these values are enclosed in the 1$\sigma$ (green) region 
of the present analysis.

\begin{figure}[t]
\centerline{\includegraphics[width=0.45\textwidth, trim=0mm 0mm 0mm 0mm]{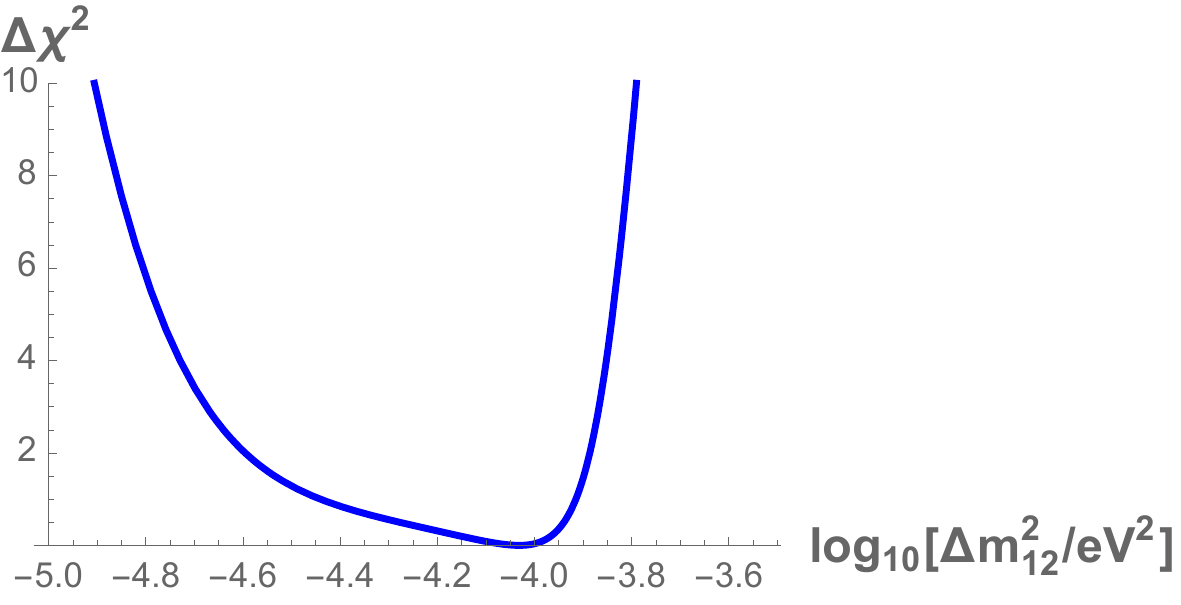}\hspace{15mm}\includegraphics[width=0.43\textwidth,trim=2mm 0.3mm 0mm 0mm]{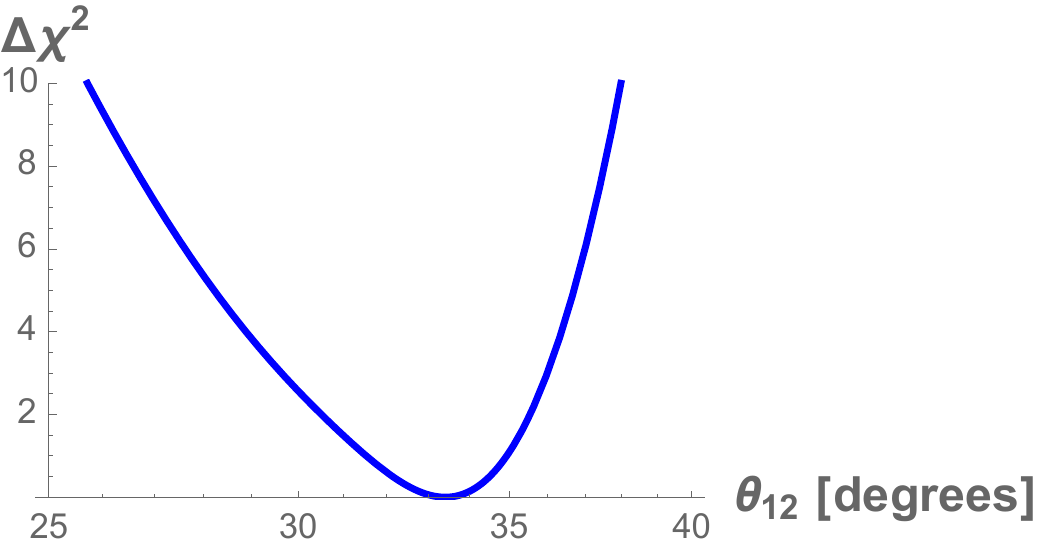}}
\caption{\em\small One-dimensional $\Delta \chi^2$ distribution, for the analysis of the solar neutrino data based on the four known values of the survival probabilities 
summarized in table~\ref{tab1}.
 \label{f33}}
\end{figure}

The one-dimensional
$\Delta \chi^2_{\mbox{\tiny 1-dof}}$ are given in Fig.~\ref{f33}. 
These curves have been 
obtained setting  $\chi^2_{\mbox{\tiny 1-dof}}=-2 \log \mathcal{L}_{\mbox{\tiny 1-dof}}$, 
where the one-dimensional likelihood 
is just the full two-dimensional likelihood, integrated 
over the other variable (i.e., a standard marginalization procedure). 
The allowed ranges, that follow from the Gaussian prescription, 
are,
\begin{equation}
\begin{array}{r|c|c}
 & \log_{10}[\Delta m^2_{21}/\mbox{eV}^2] & \theta_{12} \\[0.5ex]  \hline 
 & & \\[-2ex]
 1\sigma  \ {\scriptstyle [ \Delta\chi^2=1]}\ \  &\ \  (-4.44\ ,\ -3.91)\ \  &\ \  (31.5^\circ\ , \  34.9^\circ) \ \ \\[1ex] 
 2\sigma  \ {\scriptstyle [\Delta\chi^2=4]}\ \  &\ \  (-4.73\ ,\ -3.85) \ \   &\ \  (28.9^\circ \ , \  36.4^\circ) \ \ \\ 
\end{array}
\end{equation}
The above ranges are compatible with those given by the global fits.
In order to discuss better the meaning of these findings, 
let us consider the extremal $\Delta m^2_{21}$ values admitted at $2\sigma$ and 
let us examine the 
position of the transition region between the vacuum and the MSW regime: 
 for the lowest values, the ${}^7$Be neutrinos fall in 
in the transition region; instead, for the highest values, 
the ${}^8$B neutrinos fall in the transition region.   
This remark makes it evident that the above ranges are 
quite wide. 

It is worthwhile to repeat that 
the best fit value of $\Delta m^2_{21}$ 
of KamLAND data
is very close to
the best fit range shown above, while the value of $\Delta m^2_{21}$ that gives an optimal fit to 
the Super-Kamiokande observations  lies in the lowest border of the 1$\sigma$ region. Therefore, Borexino's data have 
some interest for the current discussion of solar neutrino findings and they indicate new ways to proceed further in the understanding of solar neutrino oscillations.

\section{Reconstruction of the survival probability\label{sec:rec}}
The likelihood $\mathcal{L}(\Delta m^2_{21},\theta_{12})$
can be then used for various purposes, 
and in particular to reconstruct statistically the survival probability at energies different
from the ones where its value is known already--i.e., to perform interpolation and extrapolation.

\begin{figure}[t]
\begin{minipage}[c]{9cm}
\includegraphics[width=8.5cm, trim=2mm -6mm 0mm 0mm]{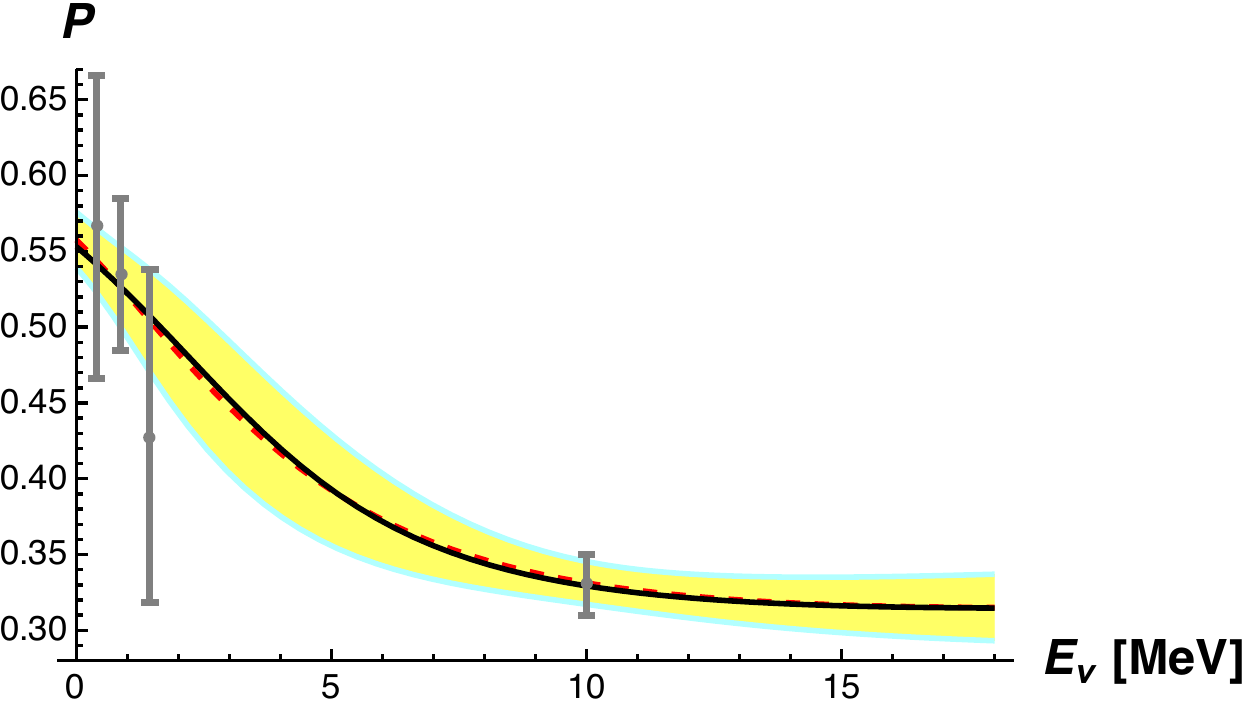}
\includegraphics[width=8.5cm, trim=0mm 6mm 0mm 0mm]{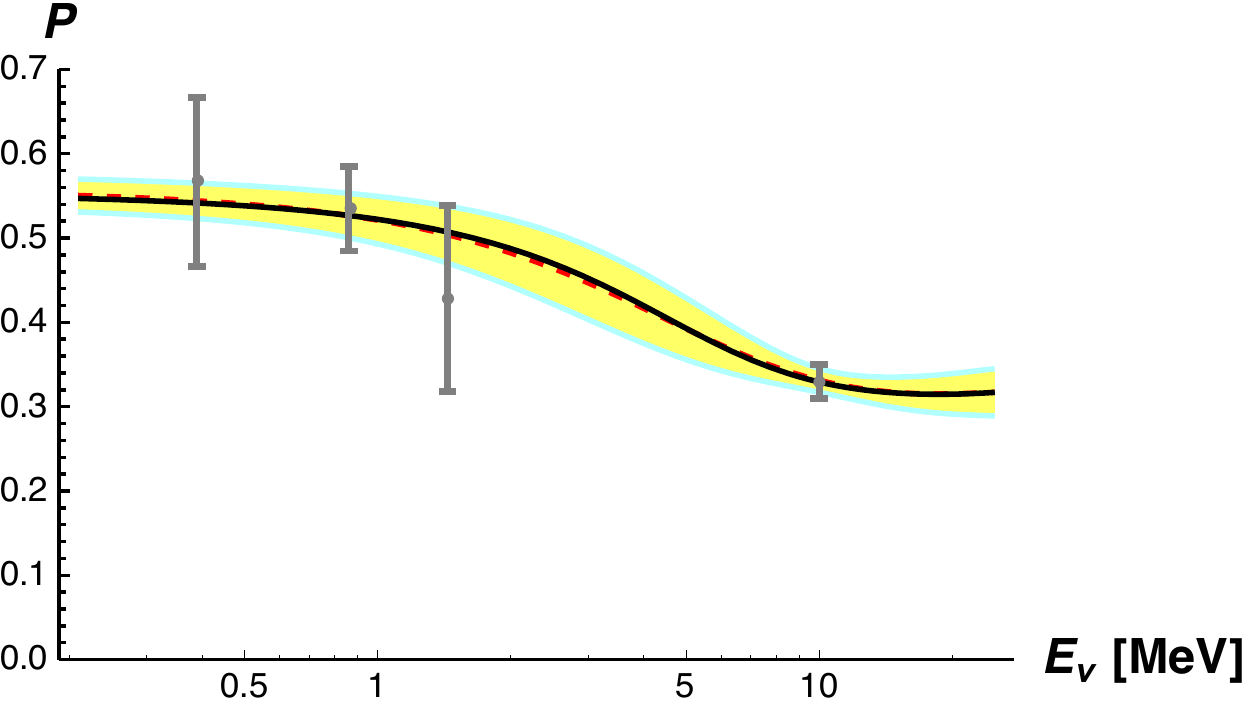}
\end{minipage}
\begin{minipage}[c]{6cm}
\caption{\em\small The survival probability of ${}^8$B neutrinos 
reconstructed from the known values, that have been obtained from the 
measured fluxes and with the help of SSM.\medskip  \newline
The yellow areas enclose the 1$\sigma$ region.
The data and the error-bars included are indicated by gray bars. \smallskip\newline
The average probability $\langle P(E_\nu)  \rangle $ (red dashed line)
and the survival probability  
$P(E_\nu) $ calculated for the best fit oscillation parameters 
of the present analysis (black continuous line) are also shown. \bigskip \newline
Top panel: plot in linear scale.
Bottom panel: plot in logarithmic scale.
\label{f4}}
\end{minipage}
\end{figure}

The most direct approach is to treat, for any value of the energy, the value of the survival probability as a random variable. 
Therefore, one evaluates the functions,
\begin{equation}
\begin{array}{c}
\langle P^a(E_\nu)\rangle \equiv \int P^a(E_\nu\, ;\, \Delta ,\theta)\times \mathcal{L}(\Delta ,\theta)\ 
d\Delta\, d\theta 
 \ \ \mbox{ with }a=1,2\ ; \\[2ex]
 \delta P(E_\nu)  \equiv \left[ \,  \langle P^2(E_\nu)\rangle -\langle P(E_\nu)\rangle^2 \, \right]^{1/2}
\end{array}
\end{equation}
thereby obtaining, for each neutrino energy $E_\nu$, the average value and 
the range of the survival probability, that are compatible with the 
dataset considered. This outcome can be then compared, e.g., with the probability 
$P(E_\nu\, ;\, \Delta m^2_{21} ,\theta_{12})$
calculated at the best fit values for  $\Delta m^2_{21}$ and~$\theta_{12}$.

The resulting survival probability 
is  shown in Fig.~\ref{f4}, using, for the two panels,   
linear and logarithmic scales. 
The plot in linear scale 
can be compared directly with Fig.~\ref{f0} and 
emphasizes the difference between low- and high-energy measurements.
The plot in logarithmic scale, instead, is often preferred 
in presentations of the data, e.g., \cite{borex}.

It is evident that the result of the procedure compares very satisfactorily with the known values of the survival probabilities 
(indicated by the vertical error-bars in gray) and that the survival probability is better constrained close  
to those energies where they are known, being more uncertain far from them.

In principle, a substantial improvement of the 
{\em theoretical} value of the beryllium  line, and of the 
{\em experimental} measurement of the $pp$ or of the 
$pep$ neutrinos, could have a big impact for the 
reconstruction of the survival probability: see again 
table~\ref{tab1} and the discussion therein for 
an assessment of the dominant error.

Before concluding, let us stress that 
Fig.~\ref{f4} shows 
the survival probability {\em of the ${}^8$B neutrinos.}
Therefore, 
for consistency, 
the three known, central values of the survival probabilities at low energies, shown in the figures by the leftmost 
grey points, do not coincide exactly with the values  given in table~\ref{tab1}.  
In fact, they are smaller by  0.7\%, 0.4\% and 1.9\% for $pp$, ${}^8$Be and  $pep$ neutrinos respectively,  
as calculated  at the best fit point and by using the SSM--see  Sect.~\ref{lj} for discussion.

\section{Summary and discussion}
Besides the disappearance of ${}^8$B neutrinos, 
there are other relevant facts that should fit into the same picture, namely the theory  
of three-flavor, solar neutrino oscillations. These include,
\begin{itemize}
\item the parameters measured by KamLAND with antineutrinos;
\item the upturn of ${}^8$B neutrinos;
\item the day-night asymmetry as measured with ${}^8$B neutrinos; 
\item the overall shape of the survival probability.
\end{itemize}
(There are also other known facts, as the 
measurements due to Homestake \cite{hs},  SAGE \cite{sg} and Gallex/GNO \cite{gx}, 
absence of an observable day-night asymmetry at lower energy \cite{b2}, the new measurement of ${}^8$B neutrinos with
a very low threshold \cite{b3}; 
in future, perhaps, also the shape of the $pp$ neutrinos and the intensity and shape of 
CNO neutrino flux could be measured.)

To date, there is a bit of tension between the first three aspects. 
No simple way out is known 
{within the conventionally accepted physics framework: 
in principle, one may object that the first measurement concerns reactor antineutrinos and not solar 
neutrinos but the standard theory predicts that $\Delta m^2_{21}$ is the same for neutrinos and antineutrinos. Moreover, }
the shape of the reactor neutrinos does not seem to need radical revisions; the shape of ${}^8$B neutrinos may be uncertain but only within percent \cite{Bahcall:1996qv, winter}; the day-night asymmetry seems to be even less unambiguous to interpret than the rest. 

Therefore, 
in this work, we focussed on the last item of the above list, exploiting the precise measurements, obtained very recently by Borexino, of 
three branches of the $pp$-chain at low energy along with SNO measurements. 

We used a very simple and transparent procedure, that moreover is adequate for the task; indeed, the  
main limitation of this analysis is just the precision of the current knowledge of the input values of the survival probability.
We checked the stability of our findings under many types of variations, e.g., omitting $pp$ and/or $pep$
data-point, using the nominal SSM prediction for the ${}^7$Be  \cite{Vinyoles:2016djt}, etc. The only relatively major aspect is the inclusion of the 
neutral current measurement of SNO.

We showed that the existing measurements of the differential flux from 4 branches of the $pp$-chain 
allow us to obtain the oscillation parameters, whose values are in good agreement with those measured by KamLAND. We indicated how to reconstruct very directly  the overall shape of the survival probability, estimating its uncertainties.

We emphasized that the standard solar model remains important for the prediction. 
Indeed, the most precise measurement of Borexino, the beryllium  line, is also the one for which the knowledge of the 
survival probability is limited by theory and not by the rate observed by Borexino. Diminishing the current theoretical uncertainty  
can have an important impact on the current discussion. 

On the other hand, it is possible at least in principle to proceed experimentally in the measurement of the $pp$ (and partly of the $pep$, in view of the CNO neutrinos) and to obtain more precise values of the survival probability, remaining free from theoretical  limitations.

\newpage

\footnotesize
\paragraph{Acknowledgments:}

I thank G.~Ranucci for the  conversation that 
triggered the present investigation~\cite{link}. 
I am grateful to G.~Bellini and to the participants in the workshop  
{\em Recent Developments in Neutrino Physics and Astrophysics,}
LNGS and GSSI, Sep.~2017  \cite{lonk}, 
and in particular to F.~Calaprice, A.~Di Leva, A.~Ianni, C.~Mascaretti, A.~Palazzo,
A.Yu.~Smirnov and F.L.~Villante for several useful discussions. 
Finally, I would like to thank an anonymous Referee of {\em 
Nuclear Physics and Atomic Energy} for a very accurate and helpful review.


\bigskip
\centerline{\bf\large \hrulefill\ \ \  References\ \ \  \hrulefill}
\vskip3mm


\begin{multicols}{2}

\renewcommand\refname{

\end{multicols}

\renewcommand{\contentsname}{\vskip-7mm}

\bigskip
\centerline{\bf\large \hrulefill\ \ \  Contents\ \ \  \hrulefill}
\vskip3mm

\tableofcontents

\end{document}